\documentclass[twocolumn]{aastex701}

\newcommand{\vol}{\mathcal{V}}
\newcommand{\refr}{\mathcal{R}}
\usepackage{multirow}
\usepackage{amsmath}
\usepackage{hyperref}

\begin{document}

\title{Quantitative Links between Formation and Atmospheric Composition for Giant Planets}

\author[0000-0003-1728-8269]{Yayaati Chachan}
\affiliation{Department of Astronomy and Astrophysics, University of California, Santa Cruz, CA95064, USA}
\email[show]{ychachan@ucsc.edu}  

\author[0000-0001-5061-0462]{Ruth Murray-Clay}
\affiliation{Department of Astronomy and Astrophysics, University of California, Santa Cruz, CA95064, USA} 
\email{rmc@ucsc.edu}  

\author[0000-0002-9843-4354]{Jonathan J. Fortney}
\affiliation{Department of Astronomy and Astrophysics, University of California, Santa Cruz, CA95064, USA}
\email{jfortney@ucsc.edu}  

\begin{abstract}
Precise atmospheric abundances of giant exoplanets are becoming increasingly prevalent and there is a need to systematically convert these measurements into metrics that inform us about the planets' accretion history. We present a quantitative framework for relating atmospheric composition to the accretion of metals and primordial gas. This includes i) new relations that simplify and generalize the connection between atmospheric metallicity and metal mass fraction, ii) methods for inferring the source of the metals and quantifying the quantity of metals accreted via different sources, and iii) relating the derived quantities to expectations from formation models. We also derive a new estimate of the minimum disk mass needed to form a planetary system based on the amount of excess metals accreted by a planet, assuming that this excess originates from the drift and evaporation of pebbles at condensation fronts. This paper is accompanied by an open-source code that implements the methods presented herein.
\end{abstract}

\section{Introduction} 
The composition of giant planet atmospheres provides an important window into the accretionary past of the planets. The presence, and in particular the enrichment, of metals in these atmospheres is a compelling indication that the provenance of planets differs from that of stars. Much insight about the formation of giant planets can be gained by understanding how giant planets acquire their metals.  In many models of planet formation, giant planets accumulate solid and gaseous material through separate processes and the proportion of material provided by each process can vary \citep[e.g.,][]{Helled2014, Ikoma2025}. Thus, the relative abundance of elements accreted as solids and gas is determined by the solid-to-gas accretion rate.  Furthermore, in different regions of a protoplanetary disk, separated by snowlines, different molecules are in solid form, meaning that detailed abundance ratios may be a powerful signature of formation location.  This insight \citep{Oberg2011} has led to many studies of the carbon-to-oxygen (C/O) ratio in exoplanet atmospheres \citep[reviewed in][]{Madhusudhan2019, Kempton2024}, but the practical utility of this approach has been limited by large uncertainties in the differential transport of materials within protoplanetary disks \citep[e.g.,][]{Oberg2016}.  Recent observational work constraining atmospheric abundances of a larger range of elements beyond C and O \citep[e.g.,][]{Smith2024, Pelletier2025, Lothringer2025, Chachan2025a, Ruffio2026, Xuan2026, Baburaj2026, Kanumalla2026, Parker2026, Xuan_2026b} provides the opportunity to use abundance ratios to constrain planet formation histories in a substantially more robust way. The characterization of elements that reside in refractory species throughout the disk provides an especially powerful way of connecting composition to formation \citep{Lothringer2021, Turrini2021, Schneider2021, Chachan2023}. These refractory elements are tracers of solids and therefore their abundance unambiguously reveals the solid-to-gas accretion rate of giant planets, removing a major degeneracy. In addition, measurement of a single refractory element can reflect the enrichment of all refractory elements to first order because they are all accreted together via solids. Measurements of multiple refractory species in ultra hot Jupiter atmospheres supports this assumption \citep{Pelletier2023, Gandhi2023}. With this grounding, elements that sometimes reside in volatiles (such as C and O) provide additional information about the origin of metals and reveal the nature of the solids that were accreted as well as the quantity of metals accreted from the gas phase.

Atmospheric measurements of extrasolar giant planets are entering a new population-level era \citep{Kempton2024, Lothringer2026}. Given the richness of the existing and upcoming datasets, we present some guiding principles and an open source tool to build connections between the atmospheres, interiors, and formation communities. The goal is to take atmospheric measurements of different elemental abundances, and use them to calculate the mass fraction and total mass (given planet mass) of each element in the planet. We use a fully-refractory species (currently the best options are S, Fe, Mg, or Si, depending on the location and temperature of the planet) to calculate the amount of solids accreted by the planet.  Then, for a specified choice of disk composition (which affects both the amounts of other elements in the solids and the composition of the gas), we calculate the amount of metals accreted from disk gas. This enables us to trace the metals and their enrichment to distinct sources, which in turn provides empirical constraints on different accretion processes and encourages closer examination of how exactly the metals are accreted. Though a range of choices remain possible for disk composition, each choice of disk model generates predictions for the relative abundances across multiple planets in the same system and each choice of dominant disk transport physics makes predictions for relative abundances in planets across different systems.  Degeneracies remain, but our method pegged to refractory accretion is significantly more robust than results obtained solely from C/O ratios. 

We note from the outset that our formulation yields constraints on composition of the outer well-mixed convective layers of planets since their composition is likely to be reflected in the observed part of the atmosphere. This outer convective zone contains a significant fraction of a giant planet's mass and it may grow over time by expanding deeper into zones that are initially stably stratified. In this process, it may entrain more heavy elements from the bottom and make the outer envelope more metal rich \citep[e.g.,][]{Vazan2018b, Fuentes2020}. Given that a negative metallicity gradient (decreasing metallicity with radius) is much more likely than a positive gradient from a formation and stability standpoint \citep{Helled2017}, our methods would provide lower limits on the metal mass present in the planet.

In addition, the presented formalism is best suited to `hydrogen-rich atmospheres' as hydrogen serves as a reference relative to which all other elemental abundances are measured. Although there is not a single precise value at which an atmosphere would no longer be considered hydrogen-rich, a reasonable value to adopt is metal enrichment of $\sim 1000 \times$ that of the star by number relative to hydrogen. For solar composition, this threshold roughly corresponds to half of all the atomic constituents being metals and a metal mass fraction of 0.934. Beyond this metallicity, it would be more appropriate to employ another abundant element as the baseline. In solar system literature, silicon is often used for this purpose, especially for meteorites, terrestrial planets, and relatively dry bodies \citep[e.g.,][]{Lodders2025}. Alternatively, water (oxygen) is utilized as a baseline for reporting the composition of comets and other icy bodies from the outer solar system \citep[e.g.,][]{Altwegg2019}.

We present the theoretical framework in section~\ref{sec:theory}, which includes methods for linking atmospheric metallicity to mass fraction, relating composition to planetary building blocks, and estimating the metal mass accreted through different sources. The applications of these methods to planets in the HR 8799 and AF Lep systems are discussed in Section~\ref{sec:examples}. In Section~\ref{sec:formation_interior}, we make connections to formation and interior models and present a novel constraint on the minimum disk mass needed to form a planet based on the amount of metals accreted via metal-enriched disk gas or solids. In Section~\ref{sec:conclusions}, we conclude with some thoughts on the future applicability of this framework. The methods laid out in this work are provided in the form of an open-source code repository called \texttt{accomp} at \url{https://github.com/y-chachan/accomp} \citep{Chachan2026a}. 

\section{Theoretical basics}
\label{sec:theory}

\subsection{Relation between atmospheric metallicity and metal mass fraction}
Planetary atmospheric elemental abundances are primarily reported in terms of the number fraction with respect to hydrogen (or another baseline element) of a given element normalized to solar or stellar number fraction of that element. To convert this number fraction to mass fraction, one typically assumes a mean molecular weight for the metals and H-He and then converts number fraction to mass fraction \citep{Fortney2013, Thorngren2019}. Here, we present a novel approach that is simpler and more generalizable compared to the prior method. Let $n_i$ be the number density of a given element, $n_H$ be the number density of H. The number fraction of element $i$ is given by 
\begin{equation}
    \chi_i = \frac{n_i}{n_H}.
\end{equation}
If $x_i$ is the planetary atmospheric metallicity of element a reported as $x_i \times$ stellar, then $x_i$ is related to $\chi_i$ by:
\begin{equation}\label{eq:xi}
    x_i \equiv \frac{\chi_{i, p}}{\chi_{i, \star}} = \frac{n_{i, p}}{n_{i, \star}}\frac{n_{H,\star}}{n_{H,p}},
\end{equation}
where the subscripts $p$ and $\star$ refer to the planet and star, respectively. The Sun's abundance is typically reported as the log (base 10) of each element's $n_i$, with log($n_H$) = 12 specified for H. We can convert the number fraction of a species to its mass fraction relative to H as follows:
\begin{equation}
    \frac{Z_i}{X} = \frac{A_i n_i}{A_H n_H},
\end{equation}
where $Z_i$ and $A_i$ are the mass fraction (i.e., fraction of the total mass) and atomic mass of element $i$, and $X$ and $A_H$ are the mass fraction and atomic mass of H. Given that $(Z_i / X)_\star$ is known apriori, we can directly relate the enrichment $x_i$ of an element $i$ to its mass fraction relative to H $(Z_i / X)_p$ in the planet
\begin{equation}
    (Z_i / X)_p = (Z_i / X)_\star x_i.
\end{equation}
Here and below, the subscripts $\star$ and $p$ outside the parenthesis refer to the given quantity in the star and the planet, respectively. The total mass fraction of metals $Z$ relative to the mass fraction $X$ of H is then just a summation over all elements heavier than H and He
\begin{equation}
    (Z/X)_p = \sum_i (Z_i / X)_\star x_i.
    \label{eq:Z_Xp}
\end{equation}
Consider first a planet that is uniformly enriched in all elements relative to the Sun such that $x_i = x$. Its metal mass fraction relative to H would therefore be 
\begin{equation}\label{eq:uniform}
    (Z/X)_p = (Z/X)_\odot x \simeq 0.0187 \, x,
\end{equation}
where the numerical value of $(Z/X)_\odot$ is derived by calculating the mass fractions of H, He, and heavy elements in the Sun \citep{Asplund2021}. The numerical coefficient is recognizable as similar to the $\approx$1.4\% total metal fraction by mass at solar metallicity, though note that Equation (\ref{eq:uniform}) uses ratio of the metal mass to the mass in hydrogen, not the fraction of the total mass in metals.

The fraction of the total mass consisting of hydrogen is 
\begin{equation}\label{eq:hfrac}
    X = (1-Z)X^\prime ,
\end{equation} where $X'$ is the H mass fraction of metal-free material (i.e., H and He only).  This reformulation leads to a simple relationship between $Z/X$ and $Z$
\begin{equation}\label{eq:zrel}
    \frac{Z}{X} = \frac{Z}{(1-Z) X'},
\end{equation}
where $X'$ is calculated from the adopted stellar abundance (e.g., $X' = 0.754$ from \citealt{Asplund2021}) in both the planet and the star under the assumption that H and He have not fractionated.  This assumption may break down for atmospheres that are marginally hydrogen poor \citep[e.g.,][]{Hu2015, Cherubim2024} but is expected to be robust for hydrogen-dominated atmospheres, including most or all giant planets. Plugging Equation (\ref{eq:zrel}), applied to the planet, into Equation (\ref{eq:uniform}) provides a direct relationship between the planetary atmospheric metallicity $x$ and the mass fraction $Z_p$ of metals in the planet:
\begin{equation}
    Z_p = \frac{(Z/X)_\odot x X'}{1 + (Z/X)_\odot x X'} = \frac{x}{(1 - Z_\odot)/ Z_\odot + x} \simeq \frac{x}{70 + x},
    \label{eq:Zp_x_relation}
\end{equation}
where we have used Equation (\ref{eq:hfrac}) and $Z_\odot = 0.014$ \citep{Asplund2021}. The inverse relationship is
\begin{equation}
    x = \frac{Z_p ( 1 - Z_\odot)}{Z_\odot (1 - Z_p )} \simeq 70 \, \bigg(\frac{Z_p}{1 - Z_p} \bigg).
\end{equation}
These expressions can be used to quickly estimate $x$ from $Z_p$ and vice versa. Similar arguments can be applied for the mass fraction of a given element rather than all the metals (equivalent to replacing $Z$ with $Z_i$ in equations above). For stellar compositions that differ from that of the Sun, one ought to replace $Z_\odot$ with $Z_\star$.

\begin{table}[]
    \centering
    \begin{tabular}{ccc}
        \hline
        Element & \multicolumn{2}{c}{$(Z_i / Z)_\odot$ (\%)} \\
         & \cite{Asplund2021} & \cite{Lodders2025} \\     \hline \hline
        O & 41.6 & 42.0\\
        C & 18.4 & 17.7 \\
        Ne & 12.3 & 13.0\\
        Fe & 8.56 & 8.24\\
        N & 5.03 & 6.10\\
        Si & 4.83 & 4.65\\
        Mg & 4.58  & 4.21\\ 
        S & 2.25 & 2.11\\ \hline
    \end{tabular}
    \caption{The fraction of the total metal mass contained in different elements in the Sun based on the abundance in \cite{Asplund2021} and \cite{Lodders2025}, in descending order. The elements listed here make up 97.6\% and 98.0\% of of the total mass of metals in the Sun for \cite{Asplund2021} and \cite{Lodders2025}, respectively.}
    \label{tab:Zi_Z_table}
\end{table}

If the enrichment of different elements is different, as generally expected, the more general expression in Equation~\ref{eq:Z_Xp} should be employed. It is instructive to rewrite Equation~\ref{eq:Z_Xp} by multiplying and dividing the right hand side by $(Z/X)_\star$
\begin{equation}
    (Z / X)_p =  (Z / X)_\star \, \sum_i (Z_i / Z)_\star x_i. 
\end{equation}
The term in the summation now depends only on the mass in element $i$ as a fraction of the total metal mass for the star. This is a fixed quantity for any given star. Table~\ref{tab:Zi_Z_table} lists $(Z_i / Z)_\star$ for metals that make up the vast majority of the metal mass in the Sun. Oxygen and carbon alone constitute 60\% of the total metal mass. It is useful to define a mass-fraction averaged enrichment $\bar{x}$
\begin{equation}
    \bar{x} = \sum_i (Z_i / Z)_\star x_i
    \label{eq:xbar}
\end{equation}
The bulk metallicity for the more general case of different enrichment can then be obtained by:
\begin{equation}
    Z_p = \frac{\bar{x}}{(1 - Z_\star) / Z_\star + \bar{x}}.
    \label{eq:Zp_multi_x_relation}
\end{equation}
For the Sun, the first term in the denominator is again $\sim 70$. For uniform enrichment, this expression simplifies to Equation~\ref{eq:Zp_x_relation}. 
Let us consider a specific example to show the utility of this expression. Consider a planet that is $10 \times$ solar enriched in O but solar in all the other elements. Since O contains 41.6\% of metals by mass in solar composition material (Table~\ref{tab:Zi_Z_table}, \citealt{Asplund2021} abundances), Equation~\ref{eq:xbar} yields $\bar{x} = 10 \times 0.416 + (1 - 0.416) = 4.74$ and Equation~\ref{eq:Zp_multi_x_relation} yields a planetary bulk metallicity $Z_p = 0.063$. Compare this with a planet that is uniformly enriched in all metals at the $10 \times$ solar ($\bar{x} = 10$) level with $Z_p = 0.124$. 
Another utility of this relation is that one can estimate the enrichment required for different elements to obtain the same metal mass fraction. All one has to ensure is that $\bar{x}$ is the same even though different elements may be contributing to the sum in the right hand side of Equation~\ref{eq:xbar}. For example, the O enrichment required to match the metal mass fraction for a $10 \times$ solar uniformly enriched atmosphere is $10 / (0.416) = 24 \times$ solar. Considering actual measurements of an exoplanet atmosphere, the measured median abundance of HR 8799 b from \citealt{Xuan2026} is C/H = 4.2, O/H = 4.9, N/H = 21.2, S/H = 5.2 relative to the star's abundance, assumed to be equal to solar abundance from \citealt{Asplund2009} (not shown in Table~\ref{tab:Zi_Z_table}) based on the average composition of the young stellar groups that HR 8799 is likely a member of. This implies $\bar{x} = 4.2 \times 0.177 + 4.9 \times 0.430 + 21.2 \times 0.0519 + 5.2 \times (0.0232 + 0.0968 + 0.0530 + 0.0498) = 5.09$, where we have included $(Z_i/Z)_\star$ for Fe, Mg, and Si along with S (all of which trace refractories and should have the same enrichment). For $Z_\star = 0.0134$ from \citealt{Asplund2009}, this corresponds to a median planetary metal mass fraction $Z_p = 0.065$. The planet's atmospheric metallicity therefore implies that nearly 6.5\% of the planet's mass is contained in metals.

We note that the approach outlined in this section is considerably easier to use compared to assuming a mean molecular weight for H-He and metals and using the number fraction to obtain mass fraction. Our approach also has the advantage of easily accommodating different levels of enrichment for different elements.

\subsection{Composition of planetary building blocks}
How is the composition of a planetary envelope related to its building blocks? To answer this question, it is helpful to separate elements that are accreted solely via solids (refractories) from those that are accreted both from solids and gas (volatiles). Refractory elements provide a critical link in this framework by providing an unambiguous tracer of solid accretion \citep{Lothringer2021, Chachan2023}. Which elements can be treated as refractory depends on the context of a planet's formation location. For close-in planets, species such as Fe, Mg, and Si can safely be considered refractory while the case of S is more equivocal. For giant planets at a few to tens of au, S is a useful tracer of refractories \citep{Kama2019}. 

Depending on the assumed formation location and the relative position of volatile snowlines, a part of the volatile elements will be in the solids. So the accretion of solids will bring in all the refractories but they will also deliver some volatile elements. Let us consider two elements, a refractory $\refr$ and a volatile $\vol$ species, for demonstration. Note that for our purposes, an element is considered ``volatile" if a significant fraction of the atom is present in volatile molecules.  For example, O is counted as volatile even though silicates contain substantial oxygen because a large fraction of O is found in H$_2$O, CO, and CO$_2$.  The calculations below can easily be generalized to more elements. Let $x_{\refr}$ and $x_{\vol}$ be their respective enrichment in a planet relative to that of the host star. Finally, let $f_{i, s}$ be the fraction of an element $i$'s total abundance that is present in disk solids with the rest ($1 - f_{i, s}$) in disk gas. This fraction is calculated under the assumption that the total amount of each species is given by the overall disk abundance, scaled to the abundance of hydrogen.  In practice, stellar abundances are used to approximate the average abundances in the disk.  In the static picture where metals in solids and gas are not radially redistributed, $f_{\refr, s} = 1$ for the refractory species and $0 \leq f_{\vol, s} \leq 1$ for the volatile species, depending on the formation location and thermal processing of solids. The volatile and refractory enrichment are therefore related by
\begin{equation}
    x_{\vol} = x_{\refr} f_{\vol, s} + (1 - f_{\vol, s}).
    \label{eq:vol_ref_relation}
\end{equation}
For example, for O, $f_{\vol, s} \approx 1$ outside the CO snowline, $f_{\vol, s} \approx 0.25$ inside the water snowline (primarily in silicates), with intermediate numbers in between.  Interior to the water snowline, the enhancement of O, $x_{\vol}$, is the sum of $x_{\vol, s} = x_{\refr}f_{\vol,s}$ from silicates and $x_{\vol, g} = (1-f_{\vol,s})$ from H$_2$O, CO, and CO$_2$ gas. We note that in this example, more refractory enhancement leads to proportionally more provision of O in silicates, but no similar factor is provided in the gas enhancement term.  This is because disk hydrogen is assumed to accrete along with the volatiles and abundances in the final planet atmosphere are measured with respect to hydrogen.

However, disks are not static and metals get redistributed due to pebble drift and evaporation, viscous evolution, and disk mass loss \citep{Oberg2016, Booth2017}. To account for these effects, we need an additional parameter that is controlled by these processes. If the disk gas has an enrichment that differs from $1 - f_{\vol, s}$, then $g_{\vol, \rm excess}$ can capture this difference. Alternatively, processes such as outward diffusion of vapor at snowlines could deposit volatiles on the solids beyond the snowline and change the volatile content of the solids to $f_{\vol, s} + s_{\vol, \rm excess}$. Equation~\ref{eq:vol_ref_relation} is then modified as follows
\begin{equation}\label{eq:vex}
    x_{\vol} = x_{\refr} (f_{\vol, s} + s_{\vol, \rm excess}) + (1 - f_{\vol, s}) + g_{\vol, \rm excess}.
\end{equation}
For elements distributed among multiple volatile molecules, excess terms of both types in Equation~\ref{eq:vex} may be present. However, to keep the number of free parameters in check while making compositional inferences, it is reasonable to only use one of these excess terms depending on the context of the planetary system. In this work and the accompanying code, we specify the enhancement in volatile abundance interior or exterior to its snowline relative to the stellar elemental abundance of one of the elements present in the volatile. For example, specifying that the disk gas is $n \times$ stellar enriched in CO inside the CO snowline and that the reference element is C implies that the excess C/H (i.e., not including local gas phase C computed under the assumption of no redistribution) in the disk gas is $n \times$ stellar\footnote{Note that because this enhancement refers only to the excess, in this context $0\times$ stellar corresponds to no enrichment and $1\times$ stellar is enriched.}. This sets $g_{\vol,\rm excess} = n$ for C. Recall that $g_{\vol,\rm excess}$ (and similarly $s_{\vol,\rm excess}$) is the ratio (by number) of the excess element to hydrogen expressed as a fraction of this ratio for the star. So though CO enrichment adds the same number of C and O atoms, it does not produce the same $g_{\vol,\rm excess}$ for both. The corresponding $g_{\vol,\rm excess}$ for O in the above example would be $n \times {\rm (C/O)}_\star$. In inner regions of disks where multiple volatile molecules might have sublimated and enriched the disk gas, we can specify the enrichment and reference element for each molecule and $g_{\vol,\rm excess}$ or $s_{\vol,\rm excess}$ would then be determined by the sum of all the different volatile enrichments that contribute to any given element.  

These expressions show that these compositional relations are identical for stars of different compositions as long as the fractional distribution of volatile $f_{\vol, s}$ is not dependent on stellar composition. For a more detailed discussion of $f_{\vol, s}$ and the values it takes for oxygen and carbon in different parts of the disk, we refer the reader to \cite{Chachan2023}.

\subsection{Estimating metal mass accreted from solids and gas}

We can use the relations developed above to relate metal mass fraction and enrichment as well as the quantify the amount of metals accumulated via different sources. Continuing with the $\refr$ and $\vol$ species considered above, the total mass fraction-averaged enrichment $\bar{x}$ is
\begin{equation}
    \bar{x} = x_{\refr} (Z_{\refr} / Z)_\star + x_{\vol} (Z_{\vol} / Z)_\star,
\end{equation}
where $(Z_{\refr} / Z)_\star$ and $(Z_{\vol} / Z)_\star$ refer to the mass in refractory and volatile species respectively as a fraction of the total metal mass in the star. Equation~\ref{eq:Zp_multi_x_relation} can then be used to calculate the metal mass fraction $Z_p$ from $\bar{x}$ of the planet.

To estimate the amount of metals accreted from different sources (e.g., solids, gas), we calculate the mass-averaged enrichment $\bar{x}_{[j]}$ from a particular source $j$. We first limit ourselves to the case of a static disk in which all the elements locally add up to stellar values. For elements accreted through solids, the mass-averaged enrichment $\bar{x}_s$ is given by
\begin{equation}
    \bar{x}_s = x_{\refr} [(Z_{\refr} / Z)_\star + f_{\vol, s} (Z_{\vol} / Z)_\star],
    \label{eq:x_bar_solids}
\end{equation}
i.e., the sum of refractories $x_{\refr} (Z_{\refr} / Z)_\star$ and the volatiles that come along with solids $x_{\vol, s} = x_{\refr} f_{\vol, s}$. Locally, we would expect the disk gas to have an enrichment of $(1 - f_{\vol, s}) \times$ solar for the volatile $\vol$. The contribution of the local gas to the mass-fraction averaged enrichment $\bar{x}_g$ would be 
\begin{equation}
    \bar{x}_g = (1 - f_{\vol, s}) (Z_{\vol} / Z)_\star.
    \label{eq:x_bar_gas}
\end{equation}
The total contribution of the solids and the local gas to the planet's volatile enrichment is $x_{\vol, s} + 1 - f_{\vol, s} = (x_{\refr} - 1) f_{\vol, s} + 1$ (see Equation~\ref{eq:vol_ref_relation}). 

If the measured enrichment of the volatile species $x_{\vol} > (x_{\refr} - 1) f_{\vol, s} + 1$, then the planet must have accreted volatiles beyond what we would expect from a static disk picture.  In other words, the amount of solid material accreted is constrained by the planet's refractory elements and the amount of gas accreted is constrained by the planet's hydrogen, putting an upper limit on the accretion of volatile elements in the absence of disk redistribution. This excess enrichment is the difference between the measured volatile enrichment and the enrichment expected in a static disk model $x_{\vol} - [(x_{\refr} - 1) f_{\vol, s} + 1]$. 
The mass-fraction averaged enrichment from this excess volatile accretion $\bar{x}_{\rm excess}$ is
\begin{equation}
    \bar{x}_{\rm excess} = [(x_{\vol} - 1) - (x_{\refr} - 1) f_{\vol, s}] (Z_{\vol} / Z)_\star.
    \label{eq:x_excess}
\end{equation}
The excess volatile mass informs us about the metal redistribution processes in protoplanetary disks. For example, if the excess volatiles are accreted from metal-enriched gas due to pebble drift and evaporation, then $g_{\vol, \rm excess}$ tells us about the level of enrichment of this gas (see Equation~\ref{eq:vex}). Similarly, $s_{\vol, \rm excess}$ would indicate the excess enrichment of the solids if these volatiles came from solids just beyond the volatile's snowline that were enriched due to outward diffusion and condensation. 

To calculate the contribution of a given metal source $j$ to the bulk metallicity, we have to use
\begin{equation}
    Z_{p, [j]} = \frac{\bar{x}_{[j]}}{\bar{x}} Z_p,
    \label{eq:component_Zp}
\end{equation}
Here, $Z_{p, [j]}$ refers to the bulk metallicity contribution of the metal source $j$, which can be solids (Equation~\ref{eq:x_bar_solids}), local gas (Equation~\ref{eq:x_bar_gas}), or excess metals (Equation~\ref{eq:x_excess}). This ensures that the bulk metallicity contributions of all the different sources add up to the bulk metallicity corresponding to the total enrichment. We cannot use Equation~\ref{eq:Zp_multi_x_relation} to estimate the bulk metallicity contribution of a particular metal source as this relation is non-linear and does not follow the additivity relation, i.e., $Z_p(\bar x_s) + Z_p(\bar x_g) + Z_p(\bar x_{\rm excess}) \neq Z_p (\bar x_s + \bar x_g + \bar x_{\rm excess})$. Equivalently, the excess contribution can be calculated using $Z_{p, {\rm excess}} = Z_{p} - Z_{p, s} - Z_{p, g}$, where Equation~\ref{eq:Zp_multi_x_relation} is used to calculate $Z_p$, and Equation~\ref{eq:component_Zp} is used to calculate $Z_{p, s}$ and $Z_{p, g}$, yielding $Z_{p, {\rm excess}} = (\bar x_{\rm excess}/\bar x)Z_p$. Once we have the total bulk metallicity and the contributions from different sources, we can use measurements of a planet's mass to convert these metal mass fractions to absolute metal mass.

Using the atmospheric abundances of a planet to quantify the mass of metals accreted from different sources provides a powerful probe of the planet's formation and accretion history. Using the volatile and refractory enrichment, we can deduce the sources of the metals. These enrichment measurements can then be converted into the mass of metals accreted by the planet from these different sources. Having a handle on the mass of the metals enables a closer link to protoplanetary disk and formation models. Our procedure does not obviate the need for a model of the processes that can redistribute volatiles, producing $g_{\vol, \rm excess}$ and $s_{\vol, \rm excess}$ terms, but it provides a straightforward way to determine whether the data requires such redistribution and to constrain which redistribution scenarios are sufficient and/or allowed. By comparing the implied composition of a planet's building blocks with the expected solid and gas composition at the planet's location, we can also deduce whether the planet's measured composition suggests migration from its formation location. In the outer regions of protoplanetary disks, the composition of building blocks can be similar over large distances because of the slow decline in temperature with orbital distance. However, migration is more likely to be important for the interpretation of composition of close-in giant planets.

\section{Example applications}
\label{sec:examples}

The above framework has been applied to observations from two directly-imaged planetary systems thus far---the four HR 8799 planets \citep{Xuan2026} and AF Lep b \citep{Xuan_2026b} with interesting results.  In each of these works, atmospheric retrievals yield the abundances of elements relative to those of their host star (i.e., the $x_i$ values for different elements as defined in Equation \ref{eq:xi}).  In \citep{Xuan_2026b} and \citep{Xuan2026}, stellar abundances were assumed to be those of the Sun because the stars' moving groups have abundances compatible with solar composition.

Both AF Lep b and the four directly imaged giant planets orbiting HR 8799 are distant enough from their host stars that sulfur may be treated as refractory (\cite{Kama2019}, see \citealt{Xuan_2026b, Xuan2026} for further discussion), so we take $x_{\refr}$ in Equation (\ref{eq:vol_ref_relation}) to be the abundance of S.  Abundances, $x_{\vol}$, of two volatile species are available for all of these planets: C and O. HR 8799 b also has a measured abundance of N.  We first test the feasibility of the simplest type of disk model under consideration---one in which no redistribution of volatiles occurs.  We solve for $f_{\vol,s}$ for each of C and O using Equation (\ref{eq:vol_ref_relation}), where $x_{\vol}$ is the measured abundance of C or O and $x_{\refr}$ is the measured abundance of S.  For AF Lep b, we find that this simple model is sufficient.  For O, $f_{\vol,s} = 0.73^{+0.22}_{-0.16}$, compatible with allocation of $\sim$75\% of the element in solid silicates and H$_2$O ice and $\sim $25\% in gaseous CO, and CO$_2$.  This allocation of oxygen is compatible with disk models at the planet's orbital distance of 9 au that partition molecular components into solids and gas without redistribution \citep[e.g.,][]{Chachan2023}.  For C, $f_{\vol,s} = 0.51^{+0.18}_{-0.13}$.  A non-redistributive disk model thus requires that a substantial portion of the refractory carbon stays in the solids at the planet's location.  The disk location at which refractory organics, known to host about half of the carbon in the interstellar medium \citep{Mishra2015}, sublimate is not well understood, but retention of this material in solids at 9 au is compatible with current constraints \cite[e.g.][]{Gail2017, Dartois2018}.

Application of Equation (\ref{eq:vol_ref_relation}) to the three inner HR 8799 planets, in contrast, yields $f_{\vol,s} > 1$ for C and O, indicating that volatile redistribution is required.  For the outermost planet, $f_{\vol,s} \sim 1$ for C and O, suggesting that these volatiles are fully condensed out into solids, and $f_{\vol,s} > 1$ for N, again indicating the need for volatile redistribution. Motivated by the location of the planets between the predicted locations of their natal disk's CO$_2$ and N$_2$ snowlines, we next apply Equation (\ref{eq:vex}) with a non-zero value for $g_{\vol,\rm excess}$.  We set $s_{\vol, \rm excess} = 0$.  Disk models indicate that pebbles may drift inward across snowlines, where their volatile material sublimates, enriching gas within the snowline locations \citep[e.g.,][]{Oberg2016, Booth2017}.  Spatially resolved observational evidence for enrichment of disk gas in CO inside the CO snowline in disks remains equivocal. For three disks around Sun-like stars, \citep{Zhang2021} report no such enrichment. For two disks around A stars (HD 163296 and MWC 480), observations suggest inner disks are enriched in CO but that this enrichment is likely centrally-peaked rather than just inside the CO snowline \citep{Bosman2021, Armitage2026}. However, these conclusions depend on comparisons with detailed disk chemical models and differences in model assumptions can significantly alter the inferred CO gas phase abundance \citep{Pascucci2023}. Nonetheless, \emph{JWST} observations of inner disks reveal water rich disk gas that broadly support the importance of radial pebble drift and evaporation in redistributing volatiles in protoplanetary disks \citep{Banzatti2023, Romero-Mirza2024}. We note that HR 8799 is an A star, like those disks for which CO enhancement interior to the snowline has been observed.  If typical disks orbiting G- and A-stars experience different amounts of CO redistribution through pebble drift, our model applied to future planetary atmosphere observations will provide a method to probe this difference, complementary to direct observations of disk gas. 

In the case of HR 8799, the most pertinent snowlines are those for CO and N$_2$. Interior to each snowline location, we require that gas enrichment $g_{\vol, \rm excess}$ occur in proportions consistent with sublimated gas.  In other words, since C and O only sublimate in the form of CO, the model's self-consistency requires that $g_{\vol, \rm excess}$ for both elements is set by the enrichment of CO in the disk gas, while the value of $g_{\vol, \rm excess}$ for N may vary independently (see note below about how N$_2$ and CO enrichment of disk gas might be related). 
For simplicity, we assume that volatile enhancement interior to each snowline is uniform in space.  This assumption may be incorrect, particularly early in the disk's lifetime when diffusion has not had the opportunity to fully mix the region.  However, we find that taking this effect into account is not needed to match the HR 8799 data.  We note that for planets located closer to their host stars, interior to several separate snowlines, a more complicated disk model is needed to determine the expected relative values for $g_{\vol, \rm excess}$ for different elements.  This complication is not required to model observed abundances in the HR 8799 planets, but models of this behavior are available in the literature \citep[e.g.,][]{Booth2017, Schneider2021, Ohno2026}. 

Interestingly, the composition of all three inner HR 8799 planets is compatible with the same level of CO enrichment of $3 \times$ solar defined using C as the reference atom (i.e., excess C/H = $3 \times$ solar), yielding $g_{\vol, \rm excess} \sim 3$ for C and  $g_{\vol, \rm excess} \sim 3 \times {\rm (C/O)_\star} \sim 1.65$ for O. This assumes that all of C and O enrichment is coming from CO given that the planets are within the CO snowline but beyond the CO$_2$ snowline.
This consistency makes a simple model consisting of CO enrichment due to pebble drift across the CO snowline very compelling. The outermost planet's N abundance puts a looser constraint $g_{\vol, \rm excess} \sim 10-30$ for N, primarily due to the effects of disequilibrium chemistry and partly due to this cold planet's dimness. The requisite enrichment of CO and N$_2$ are reasonable according to numerical disk evolution models that account for pebble drift and evaporation \citep{Booth2017, Schneider2021}. The measured composition implies that the outermost planet was likely between the CO and N$_2$ snowlines while the inner three planets were within the CO snowline \citep{Xuan2026}. Strikingly, the formation location of the planets implied by their composition is compatible with their positions relative to the estimated location of the CO and N$_2$ snowlines (55 and 80 au, respectively) in this system, suggesting that large scale migration is not required to explain their compositions. 
 
Moreover, the enrichment of CO and N$_2$ in the disk gas are expected to be different but interrelated as they depend on the same underlying total amount of solids that have drifted past the snowlines and lost their volatiles. By default, we assume that 40\% of C is contained in CO (with the rest in solids) and 60\% of N is contained in N$_2$. A $3 \times$ solar enrichment of CO in the disk gas would therefore suggest an N enrichment of $0.6 / 0.4 \times 3 = 4.5 \times$ solar under the assumption that the CO and N$_2$ snowlines are close enough together that the cumulative mass of pebbles that drifted through each is similar.  This ratio is compatible with the measurements at the $1.9~\sigma$ level. However, the exact proportion of C and N in CO and N$_2$ respectively is somewhat uncertain and for a larger amount of N is in N$_2$ ($\sim 70-80$\%) and a lower amount of C is in CO ($\sim 20$\%), the inferred disk gas enrichment would be compatible with a single value of the integrated pebble mass that has drifted past the CO and N$_2$ snowlines. A more precise measurement of N abundance in the outermost planet as well as N measurements for the inner three planets would allow a powerful test of the pebble drift and evaporation hypothesis.

\section{Connections to protoplanetary disk and formation models}
\label{sec:formation_interior}

Using the measured enrichment of metals in a planetary atmosphere, we can estimate the planet's metal mass fraction ($Z_p$, Equation~\ref{eq:Zp_multi_x_relation}) and the absolute metal mass ($Z_p \times M_p$) if abundances of a sufficient number of elements are available. Measurements of C and O are critical because these elements carry $\sim$60\% of the available metal mass in solar composition material (Table~\ref{tab:Zi_Z_table}). Knowledge of at least one refractory species is particularly useful because it is reasonable to assume that all refractories are enriched at the same level as all of them are acquired in tandem via solid accretion. Refractories  contain a large fraction ($\sim20\%$) of the metal mass and accounting for them allows us to obtain a more comprehensive estimate of the metal mass accreted by a planet. Finally, if these measurements are supplemented with characterization of N abundance, we can obtain a fairly complete census of a planet's metal repository. These estimates assume that the atmospheric metallicity is representative of a planet's bulk metallicity and the contribution of the core to the metallicity is small. This is a reasonable assumption for massive giant planets ($\gtrsim 1 \, M_{\rm J}$), especially if they have metal enriched envelopes, as the envelope can contain much more metals than a $10-15$ M$_\oplus$ core \citep{Chachan2025, Xuan2026, Xuan_2026b}. In general, we expect atmospheric metallicity to be a lower limit on the bulk metallicity. Often, all the elemental species needed to make these estimates are not available. In this case, it is useful to make some reasonable assumptions about the enrichment expected for other important elements that matter for metal mass estimation. In particular, for some of the cooler giants, only S might be observable as a tracer for refractories. Given that Fe, Mg, and Si also trace refractories that are accreted along with S via solids, we can assume that their abundance is the same as S. For unmeasured volatile elements such as Ne and sometimes N, the assumed abundance depends on the purpose of the calculation and a reasonable starting point could be to assume that $x_i = 0$ (absent from the planet) or $1$ (enriched at the stellar level) for these elements. 

We used the measured C, O, and S abundances for the HR 8799 planets and AF Lep b (and also N abundance for HR 8799 b) to place constraints on these planets' metal mass fraction and absolute metal mass \citep{Xuan2026, Xuan_2026b}. We assume $x_i = 0$ for missing volatiles to obtain a conservative estimate of the metal mass fraction. The enrichment of Fe, Mg, and Si is set equal to that of S as all of these elements are accreted via solids and their enrichment should therefore be the same.  The bulk metallicity for each planet is then calculated using Equations~\ref{eq:xbar} and \ref{eq:Zp_multi_x_relation}. The atmospheric compositions of the HR 8799 planets indicate that the planets are $3-7$\% metal by mass (the range is narrower for individual planets). Given their large masses, metal mass fractions of this range map to $70-120 ~ M_\oplus$ of metals in each planet with a cumulative total of $412^{+48}_{-43}~M_\oplus$ in all four planets. Similarly, AF Lep b has a comparable metal mass fraction of $\sim 4$\%, which translates to $56 \pm 7 ~ M_\oplus$ of metals given its slightly lower mass. These estimates hint at the enormous amount of metals contained in super-Jupiters and they are commensurate with the enriched ($3.3 \pm 0.5 \times$ relative to the host stars) bulk metallicities inferred for transiting super-Jupiters \citep{Chachan2025}, even though these populations might have formed in different regions of their protoplanetary disks. The metallicity estimates obtained from atmospheric composition also demonstrate that for this class of planets, most metals are likely contained in their envelopes and not in their $10-15~M_\oplus$ cores, in agreement with the assumption we made above.

A giant planet's bulk metallicity $Z_{p, \rm bulk}$ can also be estimated by using thermal evolution models to match the planet's observed mass, radius, and age \citep{Guillot2006, Thorngren2016, Chachan2025}. This technique works best for mature ($\gtrsim$ Gyr old) planets with precise measurements of their bulk properties. It is therefore not as successfully applicable to directly imaged planets as they tend to be young (the choice of initial entropy strongly affects bulk metallicity inferences) and their masses and radii typically have large uncertainties.
The difference between the bulk metallicity and the atmospheric metallicity is often used to estimate the core mass of a giant planet
\begin{equation}
    M_{\rm core} = (Z_{p, \rm bulk} - Z_{p, \rm atm}) \, M_{p}.
\end{equation}
In reality, estimation of $Z_{p, \rm bulk}$ from thermal evolution models assumes a certain metallicity for the atmosphere, which in turn controls the planet's thermal evolution by setting the cooling rate of the planet. Therefore, once a planet's atmospheric metallicity $Z_{p, \rm atm}$ has been measured, one should re-derive its bulk metallicity self-consistently using thermal evolution models that account for the measured atmospheric metallicity. We note that compositional gradients in the interior might lead to the deeper envelope having a larger metal mass fraction than the atmosphere. This simple estimate of the core mass therefore provides an upper limit on its value. 

Our framework allows us to go beyond a measurement of the total metal mass fraction alone. By inferring how refractory and volatile abundances are related and the sources of the metals in a planet's atmosphere (following the exercise in Section~\ref{sec:examples}), we can {\it separate} out the mass of metals accreted via solids and gas. This proves especially powerful for relating observed atmospheric composition to formation models.  The enrichment of a giant planet's envelope by solids ($x_\refr$) provides an important empirical measure of the extent of solid accretion during runaway accretion, which is hard to predict theoretically. If $x_\refr \sim 1$, then the solids might have been accreted in the form of small dust grains that were dynamically well-coupled to the gas. Enhancements in $x_\refr$ imply more efficient accretion of solids beyond this simple expectation. The absolute metal mass accreted from solids can be estimated as $M_s = Z_{p, s} M_p$, where $Z_{p, s}$ is calculated using Equations~\ref{eq:component_Zp} and \ref{eq:x_bar_solids}. With this estimate in hand, we can compare it to expectations from accretion of solids in common paradigms. 

Giant planets may accumulate solids via (1) a solid core formed prior to the onset of runaway gas accretion, (2) accumulation of solids entrained in disk gas as part of the runaway gas accretion process, or (3) accretion of solids during or after runaway accretion.  A growing solid core accumulates an atmosphere prior to the onset of runaway gas accretion, and the physics of solid accretion is modified once this atmosphere is present.  However, the total mass in such a core + atmosphere must be less than twice the solid core mass to avoid runaway gas accretion, so for rhetorical simplicity, we include solids accreted during this phase in (1).  We consider each phase in turn.

(1) Whether solids added to a giant planet in the form of what was originally a solid core can be dredged up into the convective region of the planet's envelope and contribute to the observed atmospheric metallicity remains unclear \citep{Guillot2004, Helled2022}.  As already noted, our calculations do not take into account metals that remain in a sequestered core at the time that the planet's atmosphere is observed.  We review several models for the \textit{original} accretion of a core to study the consistency of observed metal masses with pollution by core dredging.   Growing planetary cores may accrete solids in the form of planetesimals ($\sim 1 - 100$ km) that are available within the planet's feeding zone (i.e., gravitational region of influence):
\begin{equation}\label{eqn:feedzone}
    M_{\rm pls, fz} = (2 \pi r) (2 \Delta) \Sigma_{\rm pls},
\end{equation}
where $r$ is the planet's semi-major axis, $\Sigma_{\rm pls}$ is the planetesimal surface density, and $\Delta$ is the extent of the planet's feeding zone often characterized as a multiple (typically $3.5 \times R_H$, \citealt{Lissauer1993}) of the planet's Hill radius ($R_H = r (M_p/3 M_\star)^{1/3}$). 
 In the absence of gas accretion and considering planetesimal accretion only, one obtains the classic planetesimal isolation mass
\begin{equation}\label{eqn:classiciso}
    M_{\rm pls, iso} = \bigg[ 4 \pi \frac{\Delta}{R_H} \frac{1}{(3 M_\star)^{1/3}} \Sigma_{\rm pls} r^2 \bigg]^{3/2}.
\end{equation}
This isolation limit exists because the feeding zone of a growing planet increases only as $M_p^{1/3}$. 
We expect the planetesimal surface density to decline with with semi-major axis, and using the minimum mass solar nebula (MMSN, $\Sigma_{\rm pls} = 33~(r/{\rm 1 ~ au})^{-3/2}$ g cm$^{-2}$) from \cite{Chiang2010} to obtain a and $M_{\rm pls, iso}$, we get
\begin{multline}
    M_{\rm pls, iso} \approx 2.3 ~ n_{\rm MMSN}^{3/2} ~ \bigg(\frac{r}{{\rm10~au}}\bigg)^{3/4} \\ \bigg(\frac{\Delta/R_H}{3.5}\bigg)^{3/2} \bigg(\frac{M_\star}{M_\odot}\bigg)^{-1/2}  M_\oplus. 
    \label{eq:pls_iso}
\end{multline}
The factor $n_{\rm MMSN}$ modulates the planetesimal surface density relative to the MMSN value and $q = M_p/ M_\star$ is the planet-to-star mass ratio. 

Solids may also be accreted by a growing solid core in the form of mm-cm sized pebbles. Pebbles are marginally coupled to the disk gas and the effect of gas drag greatly enhances a growing planet's ability to capture them as they drift past it.  Gas-assisted growth can only operate within the planet's Hill radius, where the planet's gravity dominates over stellar tidal gravity and reduction of relative velocity between a pebble and the planet can cause the pebble to become gravitationally bound. Hence, this process is most effective for cores with Hill radii, $R_H$, smaller than the scale height of the gas disk, $H_g$, so that the full potential accretion cross section is embedded in protoplanetary disk gas.  This criterion corresponds to masses less than the thermal mass \citep[e.g.,][]{Lin1986, Chametla2021}, 
\begin{equation}
    M_{\rm th} = M_*\left(\frac{H_g}{r}\right)^3 \;\;.
\end{equation}

The accretion of pebbles is thought to be a self-limiting process for two reasons.  First, it depends on the inward radial drift of pebbles, which can significantly slow down when a planet becomes massive enough to perturb the disk gas and change its structure, though see \citet{Zhu2012, Stammler2023, Huang2025} for a contrary view. This happens roughly when the planet's mass becomes comparable to $M_{\rm th}$. \cite{Bitsch2018} provide an expression for the ``pebble isolation mass" encoding this process that includes the effect of disk turbulence parametrized by $\alpha_{\rm t}$ and pressure gradient ($\gamma =\partial$~ln~$P/\partial$~ln~$r$):
\begin{multline}
    M_{\rm pebble, iso} = 25 \bigg(\frac{M_\star}{M_\odot}\bigg) \bigg(\frac{H_{\rm g}/r}{0.05}\bigg)^3 \\ \bigg[0.34 \bigg(\frac{-3}{{\rm log_{10}(\alpha_t)}} \bigg)^4 + 0.66 \bigg]  \bigg[1 - \bigg(\frac{\gamma + 2.5}{6} \bigg) \bigg] \, M_\oplus.
    \label{eq:peb_iso}
\end{multline}

Second, pebble accretion is expected to be ``flow limited" when the small solids available for accretion are so well coupled to disk gas that they flow around the core rather than decoupling and accreting onto the planet.  Recall that during the epoch of solid core growth, even when the core has accumulated a gas atmosphere, atmospheric growth is slow and does not involve a fast inflow of gas into the bound atmosphere (circulation flows as in \citealt{Ormel2015,Cimerman2017} may be present, but the gas in these flows does not remain bound).  Solid particles fully coupled to the gas on the timescale of local gas flows do not accrete onto the planet.

The ``flow isolation mass" \citep{Rosenthal2018turb, Rosenthal2020a}, above which no available pebbles are able to accrete is 
\begin{equation}\label{eqn:flowiso}
    M_{\rm flow, iso} = M_*\left(\frac{H_g}{r}\right)^3\min\left[f^2\frac{c_s}{v_{\rm gas}}St_{\rm max}, f^{3/2}St_{\rm max}^{1/2}\right]
\end{equation}
where $v_{\rm gas}= (\eta^2v_K^2 + \alpha_t c_s^2)^{1/2}$ is the root-mean-square velocity of the disk gas relative to the local Keplerian velocity $v_K$, $\eta = -\gamma c_s^2/(2 v_K^2)$ is a measure of radial pressure support in the disk gas, $c_s$ is the local isothermal sound speed, $f$ is an order-unity constant, and $St_{\rm max} \lesssim 1$ is the Stokes number corresponding to the largest ``pebbles" available for accretion.\footnote{In the Epstein and Stokes drag regimes, the gas drag stopping time and its dimensionless equivalent, $St$, are independent of the relative velocity between the particle and the gas.  We assume this regime for calculations in this paper.  See \citet{Rosenthal2020a} for appropriate velocities in the ram pressure regime, which can dominate in the inner disk.}  The various regimes captured in Equation (\ref{eqn:flowiso}) result from the importance of the relative velocity between the growing core and disk gas at the location of the pebble, which may be dominated by radial pressure support in the disk, gas turbulence, or Keplerian shear.   

For $St_{\rm max} \sim 1$, $M_{\rm flow,iso} \sim M_{\rm pebble,iso} \sim M_{\rm th}$ to order of magnitude, though the order unity coefficients (best calibrated by numerical experiments) can differ.  Interestingly, flow isolation blocks accretion of small pebbles most effectively while pebble isolation blocks accretion of large pebbles most effectively, and these processes may work in concert to block accretion of most or all solids at masses comparable to the thermal mass, even if pebble traps at the edges of gaps opened in disks by giant planets are leaky.  The biggest difference between pebble and flow isolation, however, happens when $St_{\rm max} < 1$, either due to vigorous fragmentation of solids due to turbulence in the inner disk or where, in the outer disk, particle growth balanced by inward drift prevents growth to large Stokes numbers.  In these cases, flow isolation is expected to limit the growth of solid cores before they can reach the pebble isolation mass. In this work, we set $St_{\rm max}$ as the minimum of Stokes number set by fragmentation $St_{\rm frag} = v_{\rm frag}^2 / 3 \alpha_{\rm t} c_s^2$ and radial drift $St_{\rm drift} = \delta v_{\rm K}^2 / |\gamma| c_s^2$, where $v_{\rm frag}$ is the fragmentation velocity of dust grains and $\delta$ is the dust-to-gas mass ratio of the disk \citep{Birnstiel2012}.  These choices for the fragmentation and drift limits assume that particle relative velocities are dominated by turbulent motion, gas drag occurs in the Epstein drag regime, and $\alpha < St < 1$, all of which are broadly reasonable in outer regions of protoplanetary disks. 
(2) Very small solids that are coupled to the protoplanetary disk gas on the timescale of gas infall onto a planet undergoing runaway gas accretion may be accumulated directly by the planet as part of the gas accretion process.  In this case, dust and sufficiently small ``pebbles" are accreted in direct proportion to the amount of gas accreted---in other words, the ratio of the atmosphere's mass in accumulated solids to its mass in hydrogen is equal to the ratio (including only small solids) present in the disk.

(3) Giant planets may accrete solid planetesimals ($\sim 1 - 100$ km) during or after runaway gas accretion, as long as these solids are available within the planet's \textit{final} feeding zone \citep{Pollack1996}.  This feeding zone is set by Equation (\ref{eqn:feedzone}), applied to the planet's final mass $M_p$, which includes both solids and its massive gas envelope \citep{Thorngren2016}.  We note that the giant planet's final mass is not determined by Equation (\ref{eqn:classiciso}) because gas accretion is thought to be limited by gap opening \citep[e.g.,][]{Lissauer1993, Lissauer2009, Rosenthal2020b}.  Nevertheless, solid planetesimals large enough to dynamically decouple from the disk gas can be accreted from throughout the feeding zone in Equation (\ref{eqn:feedzone}). 

Using the same minimum mass solar nebula model applied in Equation (\ref{eq:pls_iso}, we calculate a fiducial value for the mass in planetesimals accreted from the feeding zone, $M_{\rm pls, fz}$, of  
\begin{multline}
    M_{\rm pls, fz} \approx 12 ~ n_{\rm MMSN} ~ \bigg(\frac{r}{{\rm10~au}}\bigg)^{1/2} \\ \bigg(\frac{\Delta}{3.5 \, R_H}\bigg) \bigg(\frac{q}{10^{-3}}\bigg)^{1/3}  M_\oplus. 
    \label{eq:pls_fz}
\end{multline}

Each of these baseline expectations can be compared with the amount of solid mass accreted by planets estimated through our framework. Future changes to these paradigms and additional models of solid accretion can be easily incorporated into our code and provide points of comparison between theoretical expectations and empirical estimates of solid accretion.

For AF Lep b, the measured C, O, and S abundances suggest that nearly all of the $56 \pm 7 ~ M_\oplus$ of metals were accreted via solids \citep{Xuan_2026b}. This metal mass comfortably exceeds the classical planetesimal isolation mass $\sim 1.9~M_\oplus$ ($n_{\rm MMSN} = 1$, and $M_\star = 1.2~M_\odot$), flow isolation mass $\sim 5.3~M_\oplus$ ($v_{\rm frag} = 3$ m/s, slightly evolved dust-to-gas ratio $\delta = 10^{-3}$ appropriate for $\sim 1$ Myr, fragmentation dominates for r $\lesssim 40$ au given this choice of the uncertain quantity $v_{\rm frag}$, $f = 1.75$), and pebble isolation mass $\sim 17~M_\oplus$ ($\alpha_t = 3 \times 10^{-4}$ and disk thermal structure from \citealt{Ida2016} and \citealt{Chachan2023b}, adopted for flow isolation as well) at the planet's current location of 9 au. If the planet was enriched by accretion of small solids that were entrained by accreting gas during runaway, the required enrichment of these small solids would be $4.7 \pm 0.7 \times$ stellar, i.e., the observed enrichment of refractory species represented by S. 
If instead the planet accreted these solids in the form of planetesimals at its present location, the required planetesimal surface density would be $n_{\rm MMSN} = 4 \times$ the MMSN, assuming the planet accretes everything in the feeding zone at its final mass (Equation~\ref{eq:pls_fz}). However, for single planet systems, it is harder to ascertain the planetary system architecture during formation and as a result, it is difficult to overcome degeneracies (e.g., with formation location) in these inferences. 

The true power of comparing estimated solid mass with expectations from formation models lies in multi-planet systems because they allow us to study trends in accreted solid mass as a function of semi-major axis. For HR 8799, we found that the planets show an increasing trend of accreted solid mass with semi-major axis \citep{Xuan2026}. The classic planetesimal isolation mass and flow isolation mass are too small to match the observed solid accretion. 
However, this trend is in quantitative agreement with both planetesimal accretion in a $n_{\rm MMSN} = 2 \times$ MMSN disk (Equation~\ref{eq:pls_fz}) and the expected pebble isolation mass for a disk around HR 8799 with properties typical of disks around similar stars. Given that both AF Lep and HR 8799 are more massive than the Sun and the expectation that disk mass is positively correlated with stellar mass, the inferred $n_{\rm MMSN}$ is reasonable for planetesimal surface density expected in these systems.
We note that if solid accretion for the HR 8799 planets' is due to pebble isolation, it does not entail that extremely efficient mixing of the solid core is required for us to detect the signature of this solid accretion in the envelope. Even modest mass cores ($\sim 1 ~M_\oplus$) have primordial envelopes in which pebbles ablate and deposit metals that can then be mixed up into the outer convective layer \citep{Brouwers2018}.

Observational constraints on excess metals accreted by a planet provide an additional lower limit on the disk solid mass assuming that the disk gas is enriched by evaporation of the volatile from radially drifting pebbles or disk solids are enriched due to the outward diffusion of evaporated volatile and its recondensation beyond the snowline. This calculation is similar in spirit to an equivalent calculation using the measured CO enhancement within the CO snowlines of disks \citep{Williams2025}.  If $f_{\vol, \rm [mol]}$ is the fraction of a given element contained in the molecule that evaporates at the relevant snowline, then $f_{\vol, \rm [mol]} (Z_\vol/Z)_\star$ is the fraction of a pebble's mass that evaporates at the snowline. If there are multiple elements present in the evaporating volatile, then we need to sum their contribution $\sum_{\vol} f_{\vol, \rm [mol]} (Z_\vol/Z)_\star$. The amount of metal mass in the planet that can be attributed to accretion of metal-enriched disk gas ($M_{Z, \rm{excess}}$) provides a lower limit on the total mass of volatiles that evaporated off drifting pebbles:
\begin{equation}
    M_{\rm peb-evap} = \frac{M_{Z, \rm{excess}}}{\sum_{\vol} f_{\vol, \rm [mol]} (Z_\vol/Z)_\star}
    \label{eq:peb-evap}
\end{equation}
where $M_{\rm peb-evap}$ gives the minimum mass of pebbles that must have passed through the planet's location and enriched the gas in the volatile. The metal mass $M_{Z, \rm{excess}} = Z_{p,\rm excess} \times M_p$, where $Z_{p,\rm excess} = Z_p(\bar{x}_{\rm excess}/\bar{x})$ is the mass fraction of metals accreted through metal-enriched gas (Equation~\ref{eq:component_Zp}), $\bar{x}_{\rm excess}$ is the mass fraction averaged enrichment due to excess volatile accretion (Equation~\ref{eq:x_excess}, equivalent to $g_{\vol, \rm excess} (Z_\vol/Z)_\star$ with $s_{\vol, \rm excess} = 0$), and $M_p$ is the planet mass.  
This provides an independent and typically a more conservative limit on the solid mass required compared to the existing minimum mass nebula estimates. This is because previous estimates of the minimum mass nebula use only the metal mass observed in the planetary system whereas in our formalism, the observed excess metal mass is a $< 1$ fraction of the minimum mass nebula estimate. For example, if a planet accretes CO-enriched disk gas, the CO that evaporated off drifting pebbles constituted only a fraction of their mass, and the remaining mass must still have been present even if it was not accreted onto the planet.

For the three inner HR 8799 planets, \cite{Xuan2026} estimate that the cumulative mass of excess CO accreted from enriched disk gas is $128^{+35}_{-33} \, M_\oplus$. Assuming that 40\% of C is in CO (and $40 \times {\rm (C/O)_\star} \approx 22$\% of O is in CO), $f_{\rm C, [CO]} = 0.4$ and $f_{\rm O, [CO]} = 0.22$, and the denominator of Equation~\ref{eq:peb-evap} is $= 0.4 \times 0.184 + 0.22 \times 0.416 \approx 0.17$, where $(Z_\vol/Z)_\star$ for C and O are taken from Table~\ref{tab:Zi_Z_table} based on \cite{Asplund2021}. That is, almost 17\% of the mass of a pebble is lost when it crosses the CO snowline and loses CO ice to vapor. This translates to a $M_{\rm peb-evap} = 750 \pm 200 \, M_\oplus$, which is significantly greater than the estimated metal mass ($\sim 410 \pm 45~M_\oplus$, see discussion above) of all four planets. A lower assumed fraction $f_{\rm C, [CO]}$ of C in CO would slightly decrease the numerator (because more of the observed metal mass can then be accounted for in the static disk picture) but this is more than compensated for by the decrease in the denominator, in effect  increasing $M_{\rm peb-evap}$. Similarly, the outermost planet is estimated to have accreted $21^{+18}_{-10} \, M_\oplus$ of excess N via N$_2$ enriched gas. Assuming that 60\% of N is contained in N$_2$, $f_{\rm N, [N_2]} = 0.6$, and we find that pebbles lose 3.2\% of their mass at the N$_2$ snowline. This translates to a $M_{\rm peb-evap} = 670^{+570}_{-320} \, M_\oplus$, a much looser constraint due to the less precise N abundance measurement. 

\section{Conclusions}
\label{sec:conclusions}
In this work, we established novel and simple relationships between atmospheric metallicity and metal mass fraction that are easily generalizable to stars of different compositions and varying enrichment of different elements. We then showed how the enrichment of volatile and refractory elements are related by the proportions in which a planet accretes metals from solids and gas. By relating these enrichments to the metal mass fractions, we calculate the amount of metals accreted via different sources. Estimates of the absolute amounts of metals accreted via solids and gas provide an exciting window into the assembly of planets and the properties of the disks in which they form. We discuss the application of this framework to observations of two systems, HR 8799 \citep{Xuan2026}  and AF Lep \citep{Xuan_2026b}, to demonstrate how it can utilize atmospheric composition measurements to make inferences about how giant planets are assembled.

The estimated amount of solid accretion gives us an empirical constraint on the efficiency with which solids are accreted during the runaway phase of gas giant formation. It is difficult to quantify the efficiency of solid accretion during this phase theoretically and the empirical constraints we obtain can inform future theoretical work in this area. The amount of metals accreted via gas inform us about the metal-enrichment of the disk gas, which can reveal whether or not disk evolutionary processes played an important part in setting the gas metallicity. If the required disk gas metallicity goes beyond expectations from a static disk due to, e.g., pebble drift and evaporation, then we can use the estimated excess metal mass to derive a novel estimate for the minimum disk mass for a planet's formation. The inferred enrichment of disk gas also provides constraints for theoretical disk models that are orthogonal to equivalent measurements in protoplanetary disks. 

This quantitative framework can be utilized to leverage giant planet atmospheric compositions into calculating the amount of metals accreted via different sources during the planet's formation, in particular envelope accretion. It is best suited to planets for which at least three elements with different volatilities can be measured. This could be a combination of C, O, and N \citep{Turrini2021} or it could be C and O combined with a refractory species \citep{Lothringer2021, Chachan2023}. The advantage of measuring a refractory element is that it gives us a handle on the solid-to-gas accretion rate during giant planet envelope accretion and allows us to separate the metals accreted via solids and gas. Constraints on oxygen and carbon, which are commonly obtained, are extremely useful because these elements constitute nearly 60\% of the total metal mass in Sun-like stars and their protoplanetary disks. Refractory species can be measured in the UV, optical, and near-infrared \citep{Lothringer2020b, Morley2024}. For distant giant planets at a few to tens of au, refractory species FeH, Na, and K are observable in the $1-2 \, \mu$m region \citep{Morley2024} and there is JWST program to observe these species in directly imaged planets (GO 8063, \citealt{Ruffio2025jwst}). SiO is another tracer of refractories with its fundamental band at $4 \, \mu$m and its abundance has been measured in at least one directly imaged object \citep{Parker2026}. For objects that are not hot enough to have refractories such as Fe and Si entirely in the gas phase, moderately refractory elements such as sulfur offer a promising probe of solid accretion. For transiting planets, sulfur may or may not be a tracer of solid accretion, depending on where these planets formed. More refractory species such as Fe, Si, Mg and potentially Na and K would provide better constraints for solid accretion \citep[e.g.,][]{Smith2024, Pelletier2025, Lothringer2025, Chachan2025a}. 

This framework is especially powerful in multi-planetary systems and it provides remarkable constraints on the accretion history and natal disk of the HR 8799 system \citep{Xuan2026}. Given that most observed giant planets are sans companion, it would be valuable to use this framework to understand the sources of metals during runaway accretion as well as to quantify the astrophysical scatter in the metal mass accreted via different sources across the broader population of giant planets. This exercise and a comparison of the atmospheric metallicities with the bulk metallicities of giant planets (inferred via their masses, radii, and ages) would provide a much more detailed picture of how giant planets are assembled. It would also enable us to decipher why runaway accretion is delayed beyond classical expectations and why super-Jupiter bulk metallicities flatten at super-stellar values \citep{Chachan2025}.

\section*{Acknowledgments}
We are grateful to the referee for a speedy report that helped us improve the clarity of our manuscript. All the authors acknowledge support from NASA’s Interdisciplinary Consortia for Astrobiology Research (NNH19ZDA001N-ICAR) under grant No. 80NSSC21K0597. J.J.F. acknowledges the support of NASA XRP grant 80NSSC24K0691 and an Investigator grant from the Simons Foundation.

\bibliography{sample701}{}
\bibliographystyle{aasjournal}

\end{document}